\documentclass[]{spie}  

\usepackage{amsmath,amsfonts,amssymb}
\usepackage[]{graphicx} 
\usepackage{multirow}  
\usepackage{eurosym}
\usepackage[colorlinks=true, allcolors=blue]{hyperref}

\def\CN2{\mbox{$C_N^2 $}}

\def\CT2{\mbox{$C_T^2 $}}

\def\sigmal2{\mbox{$\sigma ^{2}_{I} $}}

\title{Operational optical turbulence forecast for the Service Mode of top-class ground based telescopes}

\author[a]{Elena Masciadri}
\author[a]{Franck Lascaux}
\author[a]{Alessio Turchi}
\author[a]{Luca Fini}
\affil[a]{INAF - Osservatorio Astrofisico di Arcetri, L.go E. Fermi 5, 50125  Florence, Italy}

\authorinfo{Send correspondence to Elena Masciadri - e-mail: masciadri@arcetri.astro.it}

\begin{document} 
\maketitle

\begin{abstract}
In this contribution we present the most relevant results obtained in the context of a feasibility study (MOSE) undertaken for ESO. The principal aim of the project was to quantify the performances of an atmospherical non-hydrostatical mesoscale model (Astro-Meso-NH code) in forecasting all the main atmospherical parameters relevant for the ground-based astronomical observations and the optical turbulence (\CN2 and associated integrated astroclimatic parameters) above Cerro Paranal (site of the VLT) and Cerro Armazones (site of the E-ELT). A detailed analysis on the score of success of the predictive capacities of the system have been carried out for all the astroclimatic as well as for the atmospherical parameters. Considering the excellent results that we obtained, this study proved the opportunity to implement on these two sites an automatic system to be run nightly in an operational configuration to support the scheduling of scientific programs as well as of astronomical facilities (particularly those supported by AO systems) of the VLT and the E-ELT. At the end of 2016 a new project for the implementation of a {\it demonstrator} of an operational system to be run on the two ESO's sites will start. The fact that the system can be run simultaneously on the two sites is an ancillary appealing feature of the system. Our team is also responsible for the implementation of a similar automatic system at Mt.Graham, site of the LBT (ALTA Project). Our system/method will permit therefore to make a step ahead in the framework of the Service Mode for new generation telescopes. Among the most exciting achieved results we cite the fact that we proved to be able to forecast \CN2 profiles with a vertical resolution as high as 150 m. Such a feature is particularly crucial for all WFAO systems that require such detailed information on the OT vertical stratification on the whole 20 km above the ground. This important achievement tells us that all the WFAO systems can rely on automatic systems that are able to support their optimized use.
\end{abstract}

\keywords{optical turbulence - atmospheric effects - site testing - mesoscale modeling}

\section{INTRODUCTION}
\label{sec:intro}  

MOSE is a project co-funded by ESO and INAF. It is a feasibility study aiming to evaluate the opportunity to set-up an automatic and operational system for the forecast of the atmospheric parameters relevant for the ground-based astronomy (temperature, wind speed and direction, relative humidity) and the optical turbulence ($\CN2$ and integrated astroclimatic parameters). Among the astroclimatic parameters the seeing $\varepsilon$, the isoplanatic angle $\theta_0$ and the wavefront coherence time $\tau_0$. The project has been applied on the two most important sites of the European Southern Observatory (ESO) in the visible and infrared bands: Cerro Paranal (site of the Very Large Telescope) and Cerro Armazones (site of the European Extremely Large Telescope). The method used in our approach is the modeling, more precisely we used an atmospheric non-hydrostatic meso-scale model, called Meso-Nh (Lafore et al., 1998[\cite{lafore1998}]) for the atmospheric parameters. We used a code called Astro-Meso-Nh for the optical turbulence (Masciadri et al. 1999[\cite{masciadri1999a}]). Astro-Meso-Nh has been conceived more than a decade ago and, since there, is in continuum evolution. The optical turbulence is quantified by the $\CN2$, a 3D parameter. All the other astroclimatic parameters depends on the integral on the whole atmosphere ($\sim$ 20 km) of the $\CN2$ weighted by a weighting function that depends on the height (h), the wind speed (V) or the dynamic outer scale (L$_{0}$) at different powers:

\begin{equation}
\int_{0}^{\infty }F(h^{a},V^{b},L_{0}^{c})\cdot C_{N}^{2}dh
\end{equation}

In this contribution we briefly present the model configuration used for this study, we present the benefits that we can obtain in terms of ground-based observations particularly those supported by the adaptive optics (AO), the main results obtained so far in terms of model reliability, the conclusions and the perspectives.

\subsection{Model Configuration}

Cerro Paranal and Cerro Armazones are both located in the Chilean region. The distance between the two location is around 22 km. To carry out this study we used the grid-nesting technique that permits to perform simulations with a set of imbricated domains achieving the highest resolution in the innermost domain applied on a limited surface around the point of interest. The grid-nesting "two-way" implies a mutual interaction of the atmospheric flow between each couple of father and son domains. This guarantees a thermodynamic balanced of the atmospheric flow and it guarantees to achieve the most realistic conditions. After several tests we converged on two main configurations: the first one that we call "standard configuration" is made by three domains with the highest horizontal resolution equal to 500 m in the innermost domain. The second one, that we call "high horizontal resolution configuration", is made by five domains with the highest horizontal resolution equal to 100 m in the two innermost domains centered, respectively, on Cerro Paranal and Cerro Armazones. The latter configuration is  definitely more expensive from a computational point of view but it is necessary to well reconstruct the wind speed (particularly the strong wind speed) close to the surface. Table 1 and Table 2 (Masciadri et al. 2013[\cite{masciadri2013}]) report the number of grid points and the extensions of the domains of the two configurations. Figure 1 in Masciadri et al., 2013[\cite{masciadri2013}] and Figure 1 in Lascaux et al. 2013[\cite{lascaux2013}] show respectively the configuration at three and five domains.

\begin{figure} [ht]
\begin{center}
\begin{tabular}{cc} 
\includegraphics[height=4cm]{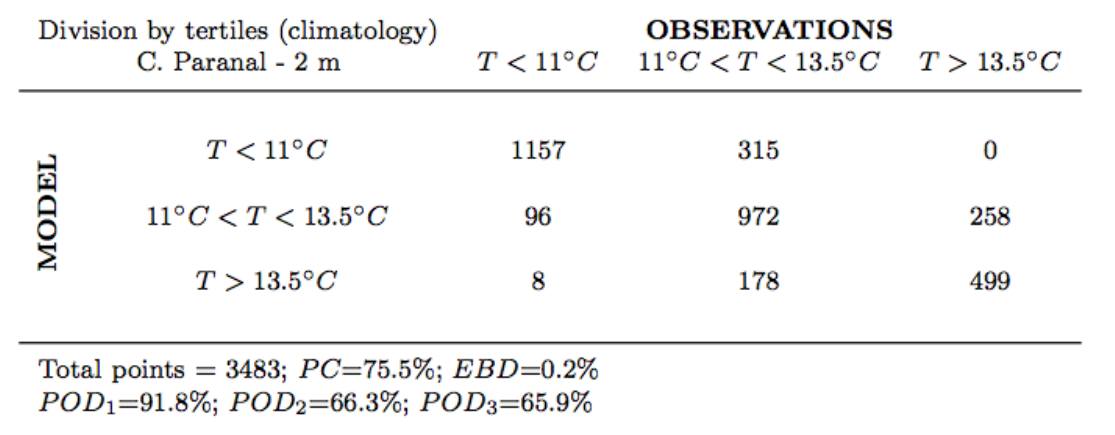}\\
\includegraphics[height=4cm]{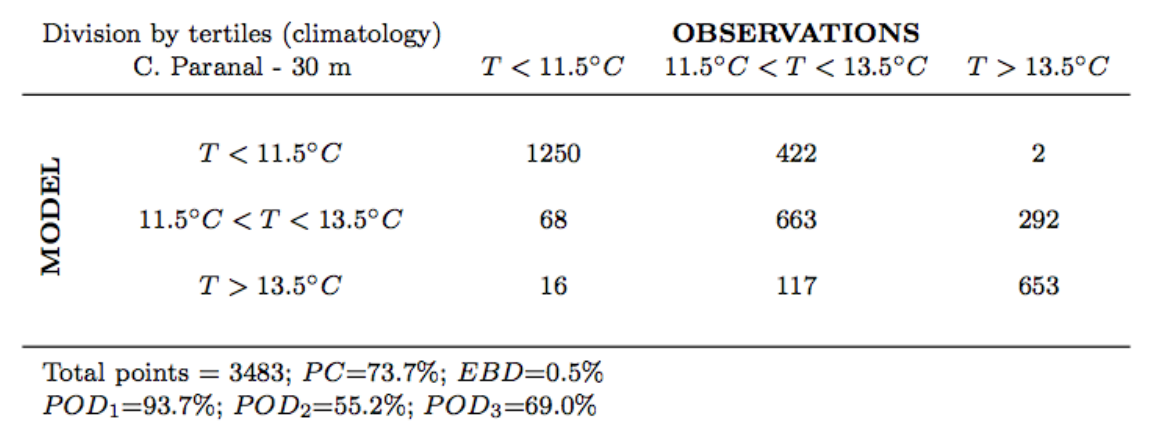}
\end{tabular}
\end{center}
\caption
{\label{fig:temp_surf} A 3$\times$3 contingency table for the temperature during the night at 2 m and 30 m above the ground at Cerro Paranal for the sample of 129 nights. Total points = 3483.}
\end{figure} 

\begin{figure} [ht]
\begin{center}
\begin{tabular}{cc} 
\includegraphics[height=4cm]{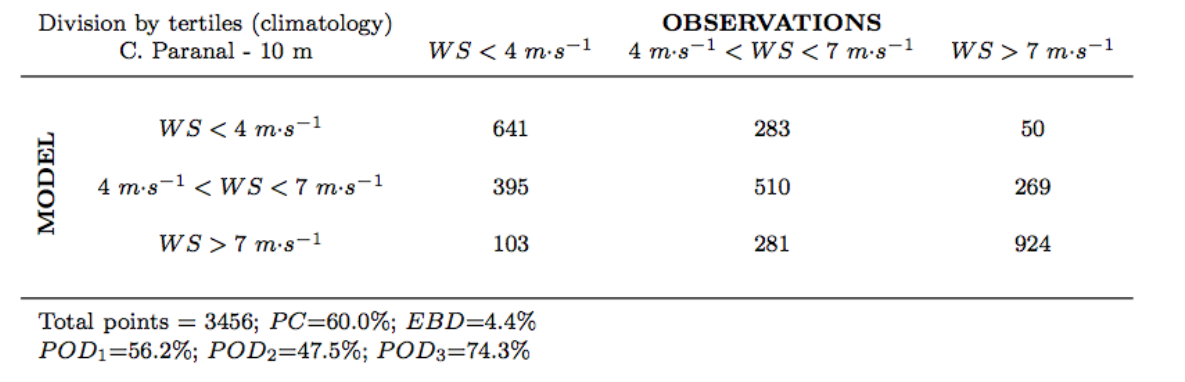}\\
\includegraphics[height=4cm]{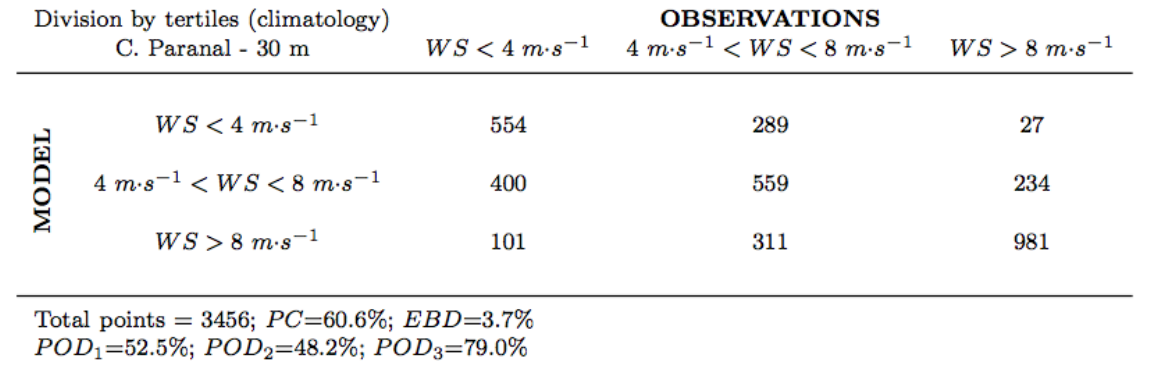}
\end{tabular}
\end{center}
\caption
{\label{fig:wind_surf} A 3$\times$3 contingency table for the wind speed during the night at 10 m and 30 m above the ground at Cerro Paranal for the sample of 129 nights. Total points = 3456.}
\end{figure} 

\section{BENEFITS IN TERMS OF ASTRONOMICAL OBSERVATIONS} 
\label{sec:ben}

Which are the benefits and feedbacks that the forecast of the atmospheric parameters and the optical turbulence can provide in terms of efficiency of observations and of management of the instrumentation ? \newline\newline
The forecast that is knowledge in advance of the temperature close to the surface is fundamental to eliminate the thermal gradient between the air inside the dome of the telescope and the primary mirror and, as a consequence, to eliminate the 'mirror seeing' that is by far the most important contribution in the total turbulent energetic budget that affects the quality of images at the focus of telescopes. Results of our analysis applied to an extended sample of 129 nights in which we compare observations (at 2 m and 30 m) with outputs of the model indicate an excellent median value of bias, RMSE and $\sigma$ not larger than 1 degree Celsius (Lascaux et al., 2015[\cite{lascaux2015}]). \newline\newline
The knowledge in advance of the wind speed close to the surface is particularly critical because is one among the main causes of vibrations of structures such as the primary mirror and the adaptive secondary mirror of telescopes. If the wind speed blows at a speed larger than a specific threshold the adaptive optics can hardly work or even not work at all. On the same rich sample of 129 nights, and observations taken at 10 m and 30 m, we obtained a median value of the bias equal to 1 ms$^{-1}$, a median value of the RMSE  equal to 2.73 ms$^{-1}$ and a median value of $\sigma$ equal to 2.69 ms$^{-1}$. If we look at the model performances applied to the single nights results are even better. We computed the bias, RMSE and $\sigma$ for each night and then we built the cumulative distribution at every level for the wind speed. In that case the median value of the RMSE is 2.06 ms$^{-1}$ and the median value of $\sigma$ is 1.25 ms$^{-1}$ (see Appendix A[\cite{lascaux2015}]). \newline\newline
The wind direction is the atmospheric parameter most easily correlated to the seeing conditions. Moreover, when the wind speed is above a certain threshold, it become extremely important to know the wind direction. Wind can indeed have a completely negligible impact on observations supported by adaptive optics or it can have a destructive impact (when the wind blows in the same direction of the line of sight). On the comparison measurements/model performed on the same sample of 129 night we found a very satisfactory model performances. If we filter out all the values of the wind speed that are weaker than 3 ms$^{-1}$ (weakly interesting from an astronomical point of view) we find excellent results: a median value of the bias equal to 3.4 degrees and a median value of the RMSE equal to 36.7 degrees that corresponds to a relative error of the RMSE$_{rel}$ (RMSE/180$^{\circ}$) equal to 20$\%$. \newline

\begin{figure} [ht]
\begin{center}
\begin{tabular}{cc} 
\includegraphics[height=6cm]{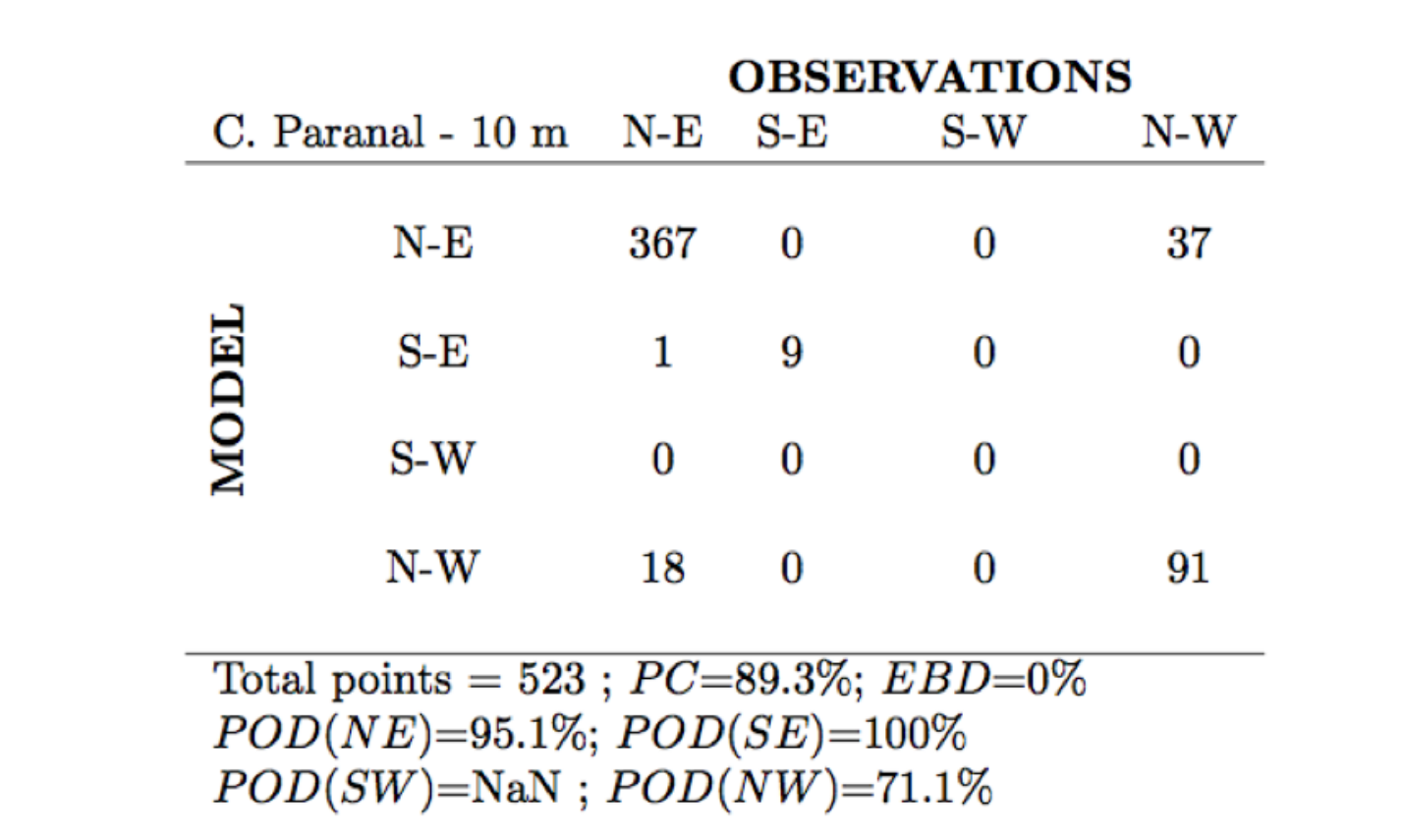}
\includegraphics[height=6cm]{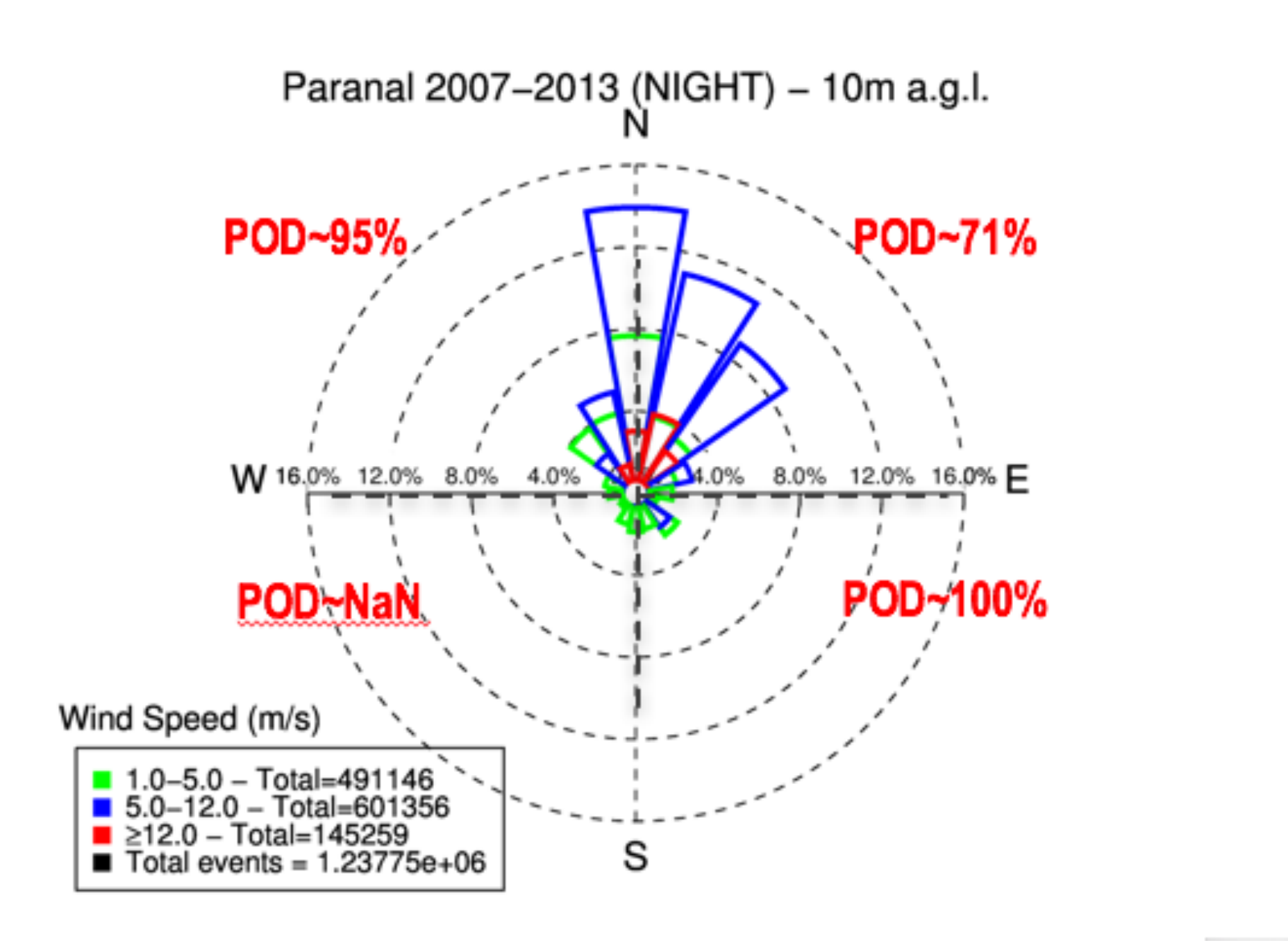}
\end{tabular}
\end{center}
\caption
{\label{fig:wind_dir} Left: contingency table of the wind direction related to those cases in which the wind speed is larger than 12 ms$^{-1}$ calculated on the sample of 129 nights. Right: wind rose of the measurements of the wind direction calculated on a sample of six years of measurements. On the same figure the values of POD$_{i}$ retrieved from  Fig.\ref{fig:wind_dir}-left side are shown.}
\end{figure} 

The Meso-Nh model presented also incredible good performances in reconstructing the vertical stratification along the 20 km above the ground of all the atmospheric parameters. We highlight results obtained for the wind speed that is one of main factor from which the calculation of the wavefront coherence time ($\tau_{0}$) depends. This parameter tells us how fast the adaptive optics (AO) system has to run. We compared the model outputs with observations (50 radiosoundings launched above Paranal in summer and winter time) and we found an excellent agreement not only in statistical terms but also comparing each radiosounding with the relative model output related to the specific radiosounding (Masciadri et al., 2013[\cite{masciadri2013}] - Appendix B). This proves that the model can follow the spatio-temporal evolution of the wind speed with a great accuracy all along the night with a very high temporal sampling. We perform indeed at present our predictions with a temporal frequency of 2 minutes but in principle there are no major constraints/limitations on the value of the temporal frequency for the model. Considering the excellent performance of the model in reconstructing the wind speed vertical stratification on the 20 km above the ground and the important advantage in terms of costs if it is compared to whatever monitor, it appears evident that this system can be very useful for applications of different nature in an Observatory, not only for the forecast finalized to the scheduling of scientific programs and instrumentation.  \newline
Indeed, due to the great reliability of the model, even if it is a model it can be considered as a reference to validate new systems for the measurements of the wind speed. We are currently collaborating with colleagues of LAM, Pontificia Universidad Catolica de Chile and Gemini to validate the principle of measurement of the wind speed used by GeMS, the multi-conjugated adaptive optics system of Gemini and first MCAO system running ok sky (Cort\'es et al., 2012[\cite{cortes2012}]). The study is not yet completed but results we obtained so far (Neichel et al., 2014[\cite{neichel2014}], Masciadri et al. 2016[\cite{masciadri2016}]) proved that the method used by GeMS can reconstruct the wind speed in a very reliable way. However this method presents two drawbacks for an operational application: (1) the maximum number of wind speed measurements provided by GeMS during a night is of 2-3 points all along the 20 km above the ground, related to 2-3 turbulent layers; (2) it is hard to figure out the automation of the measurements from GeMS currently done manually. For this reason, it seems much more realistic to simply inject the results of our model directly in the MCAO system (GeMS) for whatever use. The advantage is a complete spatio-temporal coverage all along the night of the wind speed stratification. \newline\newline
For what concerns the optical turbulence, the mesoscale models represent the unique method that is able to provide 3D maps of the $\CN2$ from which we can retrieve all the astro-climatic parameters integrated along whatever line of sight. \newline\newline
Here are reported few practical examples that indicate how the forecast of optical turbulence might impact on ground-based observations:\newline
(1) the forecast can identify the temporal windows in which the AO can not work at all (when the seeing is too strong or the $\tau_{0}$ is too weak); \newline
(2) the forecast can identify the temporal windows in which the seeing is particularly good (excellent conditions), extremely useful for scientific programs such as high contrast imaging (typical science programs are search and characterization of extra-solar planets); \newline
(3) the forecast can identify the temporal windows in which the seeing in the free atmosphere is particularly weak (challenging for observations supported by wide field adaptive optics (WFAO) that is GLAO, MCAO, LTAO, MOAO. \newline
In conclusion with just one tool we can provide a huge amount of outputs with a very important impact in different contexts of the ground-based observations.

\begin{figure} [ht]
\begin{center}
\begin{tabular}{cc} 
\includegraphics[height=4cm]{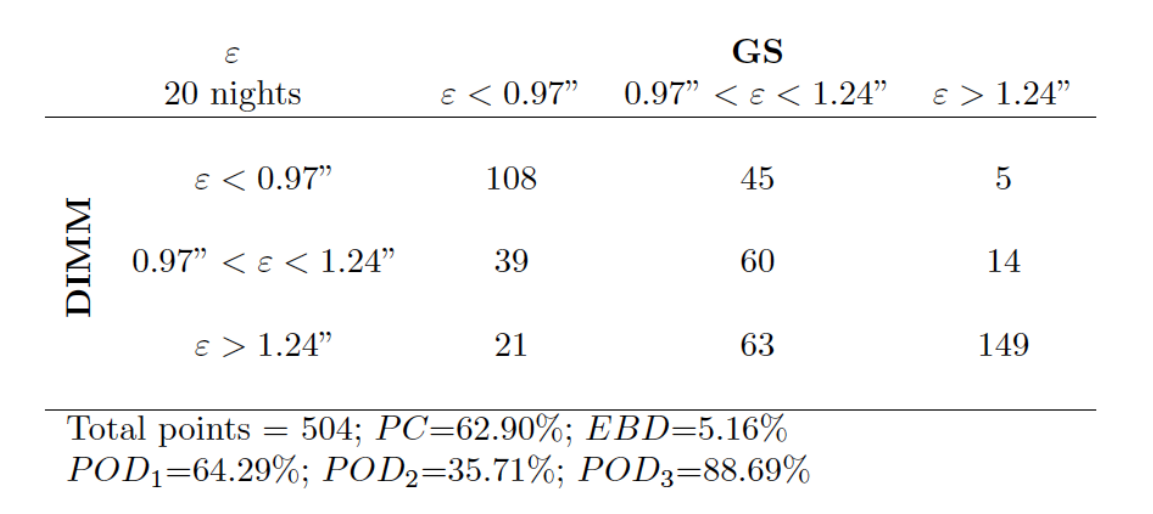}\\
\includegraphics[height=4cm]{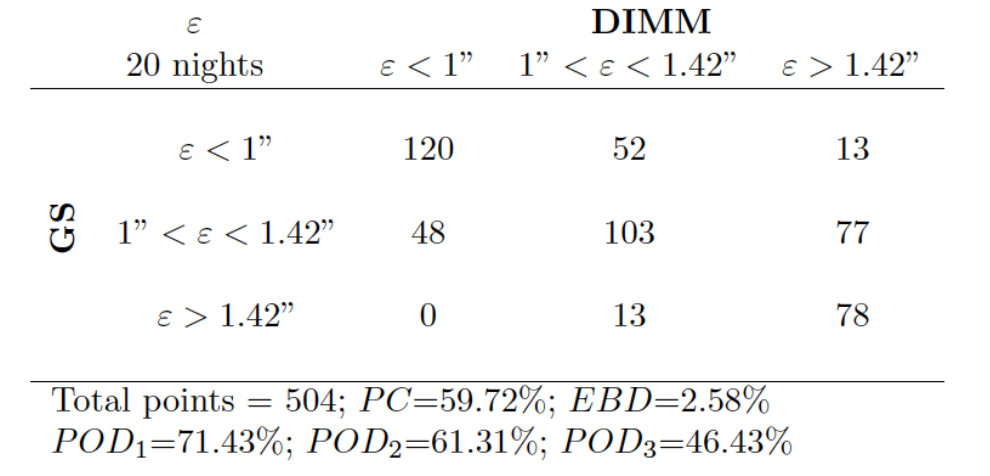}
\end{tabular}
\end{center}
\caption
{\label{fig:gs_dimm} 
A 3$\times$3 contingency table of the total seeing as measured by the DIMM and the Generalized-SCIDAR (GS). On the top GS is taken as a reference, on the bottom DIMM is taken as a reference. The sample of nights is that of the PAR2007 site testing campaign (20 nights).
}
\end{figure} 

\begin{figure} [ht]
\begin{center}
\begin{tabular}{cc} 
\includegraphics[height=4cm]{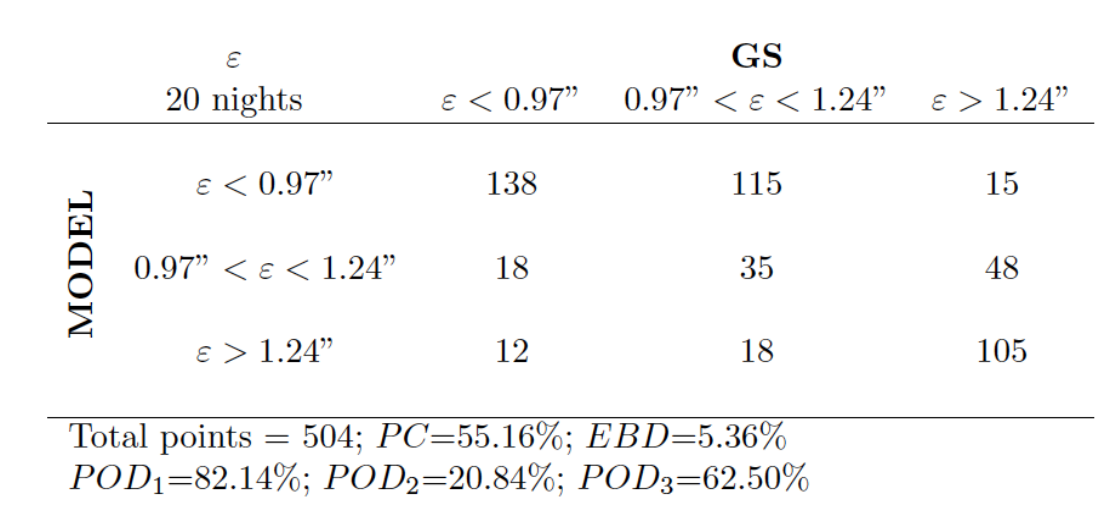}\\
\includegraphics[height=4cm]{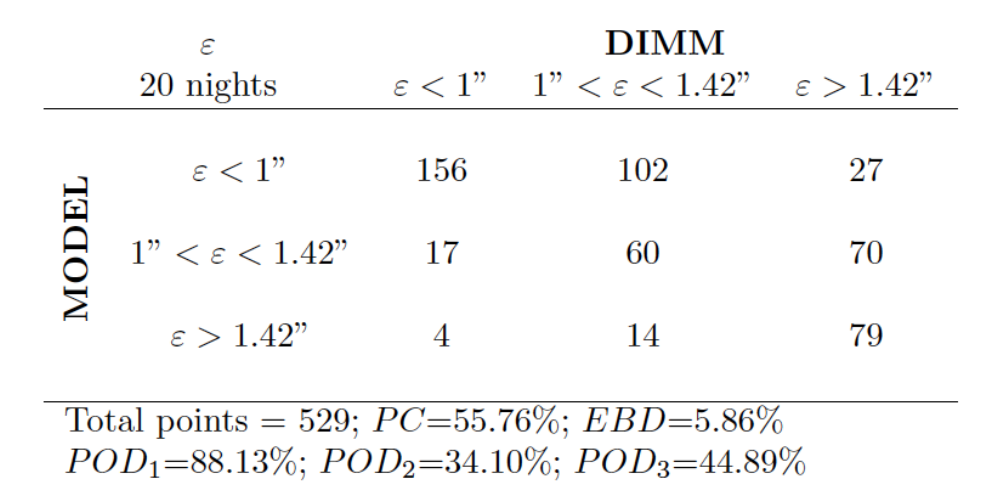}
\end{tabular}
\end{center}
\caption
{\label{fig:gs_dimm_model} 
A 3$\times$3 contingency table of the total seeing as reconstructed by the Astro-Meso-Nh and as measured by the GS (top) and as measured by the DIMM (bottom). The sample of nights is that of the PAR2007 site testing campaign (20 nights).
}
\end{figure} 

\section{CONTINGENCY TABLES}

Bias, RMSE and $\sigma$ provide fundamental information on the systematic and statistic discrepancies of the model with respect to measurements. However these parameters are not exhaustive from a practical point of view. We therefore calculated further statistical operators to have a more direct idea of the score of success of the model that implies the construction of the contingency tables. To explain briefly what a contingency table is let's assume we intend to distinguish the value of a specific variable in three categories delimited by two thresholds. A 3$\times$3 contingency table reports the number of times in which observed and simulated values belong to each category. The number of thresholds and, as a consequence, the number of the boxes N (where N$\times$N is the number of cells of the contingency table) can vary. We refer the reader to the paper of Lascaux et al., 2015[\cite{lascaux2015}] for a detailed definition of a contingency table and the statistical operators that we retrieved from the contingency tables: the percentage of correct detection (PC), the probability of detection of a parameter inside a specific interval of values (POD) and the probability of extremely bad detection (EBD). 

\section{ATMOSPHERIC PARAMETERS CLOSE TO THE SURFACE}
\label{sec:atm_surf}

We report here, as an example, the results we obtained for the temperature and the wind speed close to the surface. Fig.\ref{fig:temp_surf}  
shows the contingency tables of the temperature calculated at 2 m and 30 m. Fig.\ref{fig:wind_surf} shows the contingency tables related to the wind speed at 10 m and 30 m.  

The two thresholds (i.e. the values that determines the 3$\times$3 contingency tables) selected in both cases (temperature and wind speed) are the first and third tertiles calculated on a climatologic scale (measurements related to six years 2006-2012). By looking at Fig.\ref{fig:temp_surf} we conclude that all the POD$_{i}$ are above 66$\%$, in some cases they take values larger than 90$\%$. They are therefore well above the limit of 33$\%$ that corresponds to the random case i.e. the case in which there is no gain in using the predictive method. PC is above 74$\%$. Model performances in forecasting the temperature are therefore very promising. By looking at Fig.\ref{fig:wind_surf} we observe a more variegate set of results. The most interesting value for astronomical applications in this case is POD$_{3}$ i.e. the probability to detect a wind speed larger than the third tertiles. POD$_{3}$ is equal to 74.3$\%$ and 79$\%$ at 10 m and 30 m and this means that the model performance in detecting a strong wind speed is very good. PC is always larger than 60$\%$ that indicates a good model performances too. We refer the reader to Lascaux et al. 2015[\cite{lascaux2015}] for a more detailed analysis of the sample of 129 nights including also a seasonal variation study. In Fig.\ref{fig:wind_dir} we show an interesting case for astronomical applications. On the left is shown the contingency table (4$\times$4)\footnote{The contingency table for the wind direction is necessarily 4$\times$4 because the wind direction is distributed on 360$^{\circ}$. See Lascaux et al, 2015[\cite{lascaux2015}]} for the wind direction in which we filtered out all values of the wind speed weaker than 12 ms$^{-1}$ (we are selecting therefore the very strong wind speed).  We obtain a PC = 89$\%$ and all the POD$_{i}$ larger than 71$\%$. In Fig.\ref{fig:wind_dir}-right side, the wind rose shows the distribution of the wind direction calculated on a six years time scale together with the different POD$_{i}$. We observe that POD$_{i}$ are very large in all sectors (included those sectors in which the wind blows more frequently). The model did not register 
cases in which the wind blowed from South-West but the wind-rose tells us that wind almost never blows from that direction. Model outputs are therefore perfectly correlated to measurements exactly in those conditions that are mostly critical for astronomers i.e. when the wind speed is particularly strong.

\begin{figure} [ht]
\begin{center}
\begin{tabular}{cc} 
\includegraphics[height=15cm]{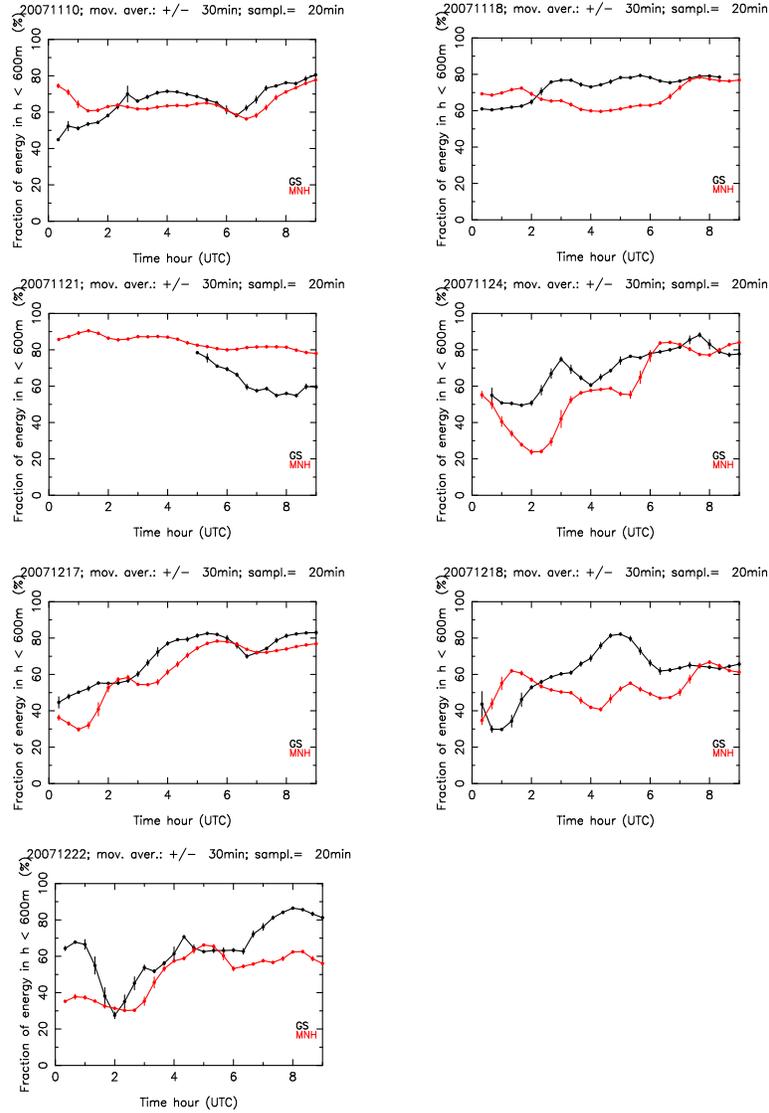}
\end{tabular}
\end{center}
\caption
{\label{fig:paper_spie_1} Temporal evolution of the fraction of turbulent energy in the first 600 m with respect to the total turbulence energy on the whole 20 km as measured by the Generalized SCIDAR and as estimated by the Astro-Meso-Nh model.}
\end{figure} 

\begin{figure} [ht]
\begin{center}
\begin{tabular}{cc} 
\includegraphics[height=6.5cm,angle=-90]{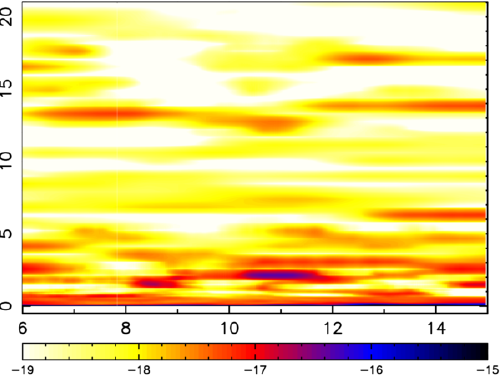}
\includegraphics[height=6.5cm,angle=-90]{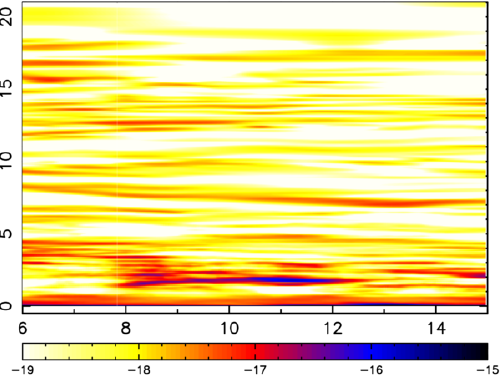}
\end{tabular}
\end{center}
\caption
{\label{fig:cn2_hr} Temporal evolution of the $\CN2$ reconstructed by the Astro-Meso-Nh model extended on around 20 km above the ground during a night at Cerro Paranal. Left: 62 vertical model levels. Right: 173 vertical model levels.}
\end{figure} 

\section{OPTICAL TURBULENCE}
\label{sec:opt_turb}

Model performances in reconstructing the astroclimatic parameters required a complex and an extended analysis. The model requires first to be calibrated using a set of measurements. It is enough to say here that MASS measurements revealed to be not sufficiently reliable (see Masciadri et al., 2014[\cite{masciadri2014}]). Measurements from a Generalized SCIDAR were therefore necessary. We used measurements related to the PAR2007 site testing campaign (Dali Ali et al., 2010[\cite{dali2010}]) successively corrected by Masciadri et al. 2012[\cite{masciadri2012}] to take into account the right normalization of the autocorrelation of the scintillation maps (as treated by Johnston et al., 2002[\cite{johnston2002}] and Avila \& Cuevas, 2009[\cite{avila2009}]).
A forthcoming paper will treat in a detailed and extensive way this analysis. Here we want to highlight a few important results that provide an idea of the Astro-Meso-Nh model performances. Fig.\ref{fig:gs_dimm} shows the contingency tables of the total seeing as measured by the Generalized-SCIDAR and the DIMM. On the top GS is taken as a reference, on the bottom DIMM is taken as a reference. The sample of nights is that of the PAR2007 site testing campaign (20 nights). Fig.\ref{fig:gs_dimm_model} shows the contingency tables of the total seeing as reconstructed by the Astro-Meso-Nh model and as measured by the GS (top) and as measured by the DIMM (bottom). The thresholds of the contingency tables are the first and third tertiles of the sample of nights we considered. It is possible to observe that the model is less performant for the seeing than for the atmospheric parameters close to the ground but two major conclusions can be retrieved:
(1) the probability of the model to detect a seeing weaker than the first tertile (POD$_{1}$ that is by far the most appealing and critical thing from an astronomical point of view) is pretty good: POD$_{1}$ = 82.14$\%$ and 88.13$\%$  if we consider as a reference respectively the GS or the DIMM; (2) POD$_{1}$ obtained between model and instruments (GS or DIMM - Fig.\ref{fig:gs_dimm_model}) is definitely comparable to POD$_{1}$ obtained between GS and DIMM (Fig.\ref{fig:gs_dimm}). This tells us that the Astro-Meso-Nh perform within the intrinsic uncertainty we have in quantifying the seeing and this also tells us that it is extremely important to have at least two independent measurements of the optical turbulence. With the term 'independent' we mean not cross-correlated among them.

Another interesting thing we can estimate with Astro-meso-Nh model is the fraction of the turbulent energetic budget contained in the first 600 m with respect to the turbulent energetic budget included in the whole 20 km (J$_{BL}$/J$_{TOT}$). Fig.\ref{fig:paper_spie_1} shows the temporal evolution of J$_{BL}$/J$_{TOT}$ during the whole night on a sample of a few nights as measured by the Generalized-SCIDAR (red line) and as reconstructed by the model (black line). This figure of merit is extremely useful for observations to be done with whatever wide field adaptive optics system (MCAO, GLAO, MOAO, LTAO). In the context the MOSE project we performed this analysis in perspective to an application to AOF at the VLT (Madec et al., 2016[\cite{madec2016}]).

In the context of the MOSE feasibility study we could also proved that the Astro-Meso-Nh model is able to reconstruct $\CN2$ profiles with a vertical resolution as high as 150 m. Fig. \ref{fig:cn2_hr} shows the temporal evolution of the $\CN2$ along one night above Cerro Paranal obtained with 62 (left) and 173 vertical model levels (right). It is evident that the number of the turbulent layers increases and the thickness of the turbulent layers becomes thinner when the number of vertical mode levels increases (Fig.\ref{fig:cn2_hr} - right). This result was possible thanks to a new algorithm of the $\CN2$ we implemented in Astro-Meso-Nh. A forthcoming paper will deal in detail this topic. This is a great achievement as such a kind of information is definitely crucial for testing the feasibility of wide field adaptive optics (WFAO) systems requiring a very high vertical resolution $\CN2$ (see the review from Masciadri et al., 2013[\cite{masciadri2013b}]). $\CN2$ profiles with such a high vertical resolution have already been used for preliminary tests on the performances of the project HARMONI (Neichel et al. 2016[\cite{neichel2016}]) and for a more extended analysis on the ultimate limitations of tomographic reconstruction for WFAO systems on E-ELT (Fusco et al., 2016[\cite{fusco2016}]).

\begin{figure} [ht]
\begin{center}
\begin{tabular}{cc} 
\includegraphics[height=8cm]{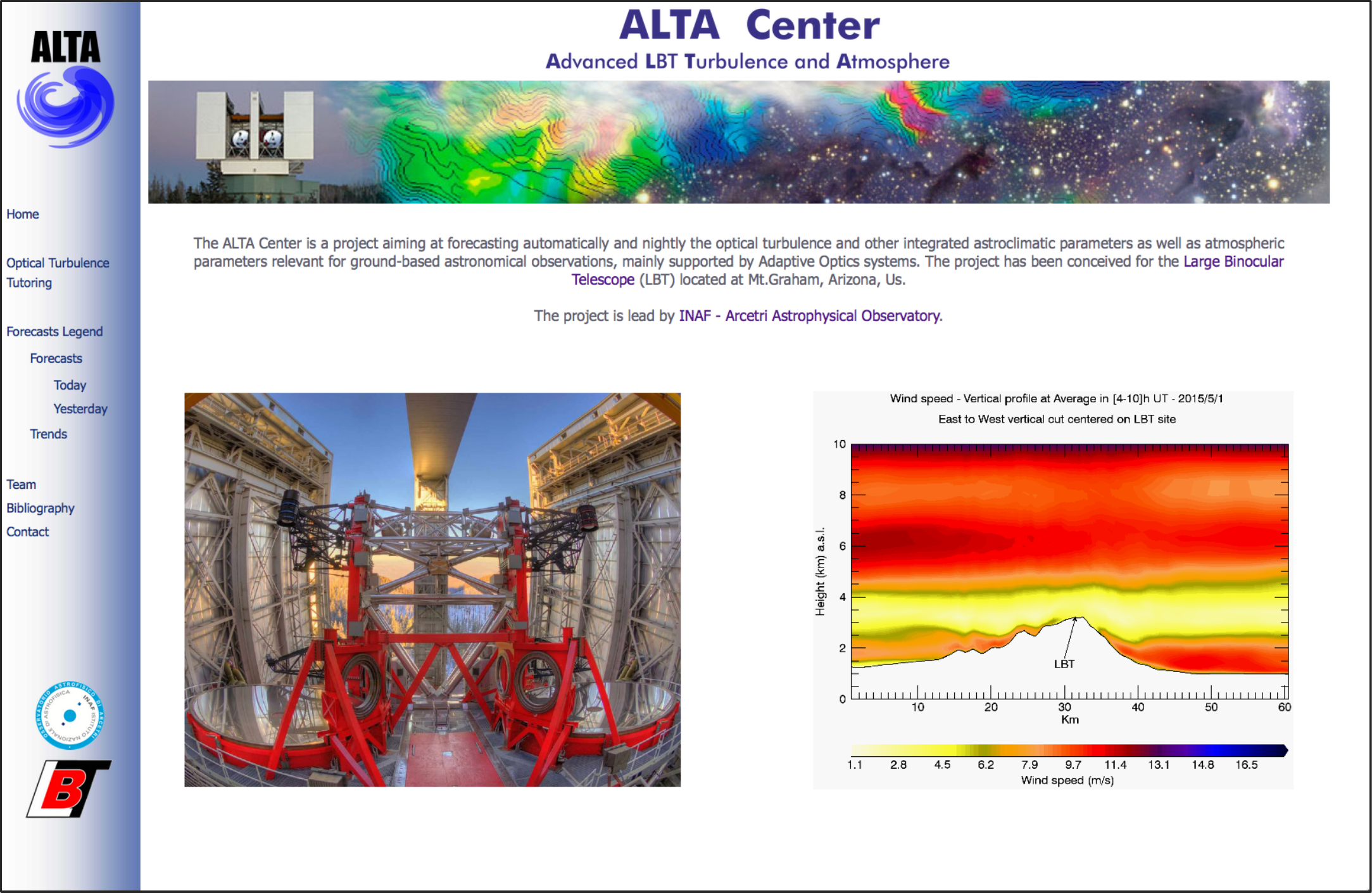}
\end{tabular}
\end{center}
\caption
{\label{fig:alta} Grab of the web page of the ALTA Center project ($http://alta.arcetri.astro.it$). ALTA is supposed to deliver forecasts of the optical turbulence, astroclimatic parameters and atmospheric parameters relevant for ground-based astronomy on a nightly time scale above Mt. Graham, site of the Large Binocular Telescope (LBT). }
\end{figure}

In terms of operational application we can say that we have already completed the automation of the whole process conceived to provide the forecasts of the optical turbulence, the astroclimatic parameters and the atmospheric parameters relevant for the ground-based astronomy on a nightly time scale. This has been done in the context of the ALTA Center (Advanced LBT Turbulence and Atmosphere Center) project. ALTA aim consists in developing an automatic system for the forecasts of the above cited parameters on Mt. Graham, site of the Large Binocular Telescope (LBT). Fig.\ref{fig:alta} shows the grab of the home of the ALTA web page, that is supposed to display the model outputs nightly. Within the end of 2016 the commissioning will be completed. This means that the model validation (for atmospheric parameters and optical turbulence) in its main baseline will be completed. Preliminary results on the model validation in reconstructing reliable atmospheric parameters in the context of ALTA have been presented by Turchi et al. 2016[\cite{turchi2016}]. A practical use of the forecasts produced by ALTA in the scheduling of the observations done at LBT has been presented by Veillet et al. 2016[\cite{veillet2016}].

\section{CONCLUSIONS}
\label{sec:conc}

We summarize here the main conclusions presented in this contribution:\\
(1) We proved that we are able to forecast the atmospheric parameters and the optical turbulence with a score of success that is already sufficiently high to definitely guarantee a concrete positive impact on the Service Mode of top-class telescopes and ELTs; \\
(2) We proved that we are able to reconstruct $\CN2$ profiles putting in evidence turbulence layers as thin as 150 m. Such a kind of information is crucial for the WFAO and open new perspectives for these sophisticated typology of AO systems; \\
(3) In October 2015 the MOSE Final Review took place successfully. In March 2016 we concluded the negotiation with ESO for the implementation of an automatic operational system that is a {\it demonstrator} of an operational version of MOSE conceived for Cerro Paranal and Cerro Armazones. Starting date, budget and duration of the this new project have been defined in agreement with the MOSE ESO Board; \\
(4) We are already involved in a similar project conceived for the Large Binocular Telescope (ALTA Center). Within the end of the 2016 we are supposed to complete the commissioning. An automatic and operational forecast system is currently running at Mt. Graham. \\ \\
The main perspectives can be summarized as it follows:\\
(1) there is still space for improving the $\CN2$ algorithm and, as a consequence, the model performances. This activity is on-going and is in a constant development in our team;\\
(2) it is important now to define the requirements of the different instruments of Paranal and Armazones to conceive a tool equivalent to ALTA Center for Paranal and Armazones. This work will be done in coordination with the MOSE ESO Board and the E-ELT Science Data Operation Division of ESO. For Paranal we will have to consider AOF, SPHERE (among those that ESO considers with maximum priority). For the E-ELT we will have to consider MAORY, HIRES, HARMONI and MOSAIC.  \\
(3)  the model calibration is sensitive to the richness of the sample of measurements. With our demonstrator we will be able to access to a much larger statistic of measurements ($\CN2$). Indeed, an intense site testing performed with a Stereo-SCIDAR is on-going at Cerro Paranal (Osborn et al., 2016[\cite{osborn2016}]). \\
(4) We are interested in extending the implemention of our tool to other top-class ground-based  facilities in addition to LBT, VLT, E-ELT including solar observatories. The most interesting sites are those equipped with instrumentation for optical turbulence measurements.

\acknowledgments 
This study has been co-funded by the ESO contract: E-SOW-ESO-245-0933 (MOSE Project).
We are very grateful to the ESO Board of MOSE (Marc Sarazin, Pierre-Yves Madec, Florian Madec and Harald Kuntschner) for their constant support to this study. Part of simulations presented in this study are run on the HPCF cluster of the European Centre for Medium Weather Forecasts (ECMWF) - Project SPITFOT. 

\bibliographystyle{spiebib} 

\end{document}